# Investigation on the Physical Properties of Two Laves Phase Compounds HRh$_2$ (H = Ca and La): A DFT Study

Md. Zahidur Rahaman[1]

*Department of Physics, Pabna University of Science and Technology, Pabna-6600, Bangladesh*
*zahidur.physics@gmail.com*

Md. Atikur Rahman[2*]

*Department of Physics, Pabna University of Science and Technology, Pabna-6600, Bangladesh*
*atik0707phy@gmail.com*

**Abstract**

Structural, elastic, electronic and optical properties of laves phase intermetallic compounds CaRh$_2$ and LaRh$_2$ prototype with MgCu$_2$ are investigated by using the first principle calculations. These calculations stand on density functional theory (DFT) from CASTEP code. The calculated lattice parameters are consistent with the experimental values. The significant elastic properties, like as bulk modulus *B*, shear modulus *G*, Young's modulus *Y* and the Poisson's ratio *v* are determined by applying the Voigt-Reuss-Hill (VRH) approximation. The analysis of Pugh's ratio shows the ductile nature of both the phases. Metallic conductivity is observed for both the compounds. Most of the contribution originates from Rh-4d states at Fermi level in DOS. The study of bonding characteristics reveals the existence of ionic and metallic bonds in both intermetallics. The study of optical properties indicates that maximum reflectivity occurs in low energy region implying the characteristics of high conductance of both the phases. Absorption quality of both the phases is good in the visible region.

---

[*]Corresponding author.

## I. Introduction

Laves phase family is one of the oldest crystal family exhibits a wide variety of physical, chemical and magnetic properties. Laves phase intermetallic compounds have corrosion and creep resistance act as magnetic materials, magneto-optical materials, hydrogen storage materials etc. [1-4]. The face centered cubic $C15$ ($MgCu_2$), hexagonal C14 ($MgZn_2$), and double-hexagonal C36 ($MgNi_2$) are the types of laves phase expressed by Fritz laves [5]. The Laves phase C15 type crystal which is favorable for superconductivity with extensively varying $T_c$ from 0.07K to above 10k [5]. As a result of high transition temperature in the similar crystal system, the detailed study on laves phases is very interesting. The clear understanding of different superconducting system is still challenge in the field of condensed matter physics. The cubic laves phase compounds are easy to study because of their simple structure and chemical composition [6, 7]. Smith et al. [8] who first investigated the electronic properties and thermal expansion character of hexagonal laves phase $CaMg_2$ and cubic laves phase $MgCu_2$. By the first principle calculations Zhang et al. analyzes the formation enthalpy of cubic laves phase $MgCu_2$ [9]. The detailed thermodynamic properties of $Al_2Ca$ and $Al_2Mg$ laves phase and mechanical properties and lattice dynamics of $ZrW_2$ and $HfW_2$ laves phase are investigated by E. Deligoz et al. [10-11]. Laves phase intermetallic compound $CaRh_2$ prototype with $MgCu_2$ was first reported in 1958 by Wood et al. [12]. They performed detailed structural characterization of $CaRh_2$ by powder diffraction. $LaRh_2$ was first reported by A.R Edwards in 1972 [13]. He investigated detailed lattice dimensions of this intermetallic. Superconductivity in $LaRh_2$ with transition temperature below 1 K was investigated by S. Pauline et al. in 1992 [14]. Except the structural characterization and superconducting nature there is no literature available yet to discuss the detailed physical properties of these two laves phase.

In this present study we decide to carry out thorough investigation on different physical properties (structural, elastic, electronic and optical properties) of these two laves phase compounds $CaRh_2$ and $LaRh_2$ by theoretical means. Thorough comparisons among the evaluated physical properties have also been done with proper discussion. The remaining part of this present work is organized as follows: The computational methods are discussed in section 2, Investigated results and the related discussion are presented in section 3 and finally the summary of the whole investigation are presented in section 4.

## II. Computational details

The present investigation were performed by using the Cambridge Serial Total Energy Package (CASTEP) code [15], based on density functional theory together with the generalized gradient approximation (GGA) with the PBE exchange-correlation functional [16-20]. For accuracy and comparison we have also used LDA (Local Density Approximation) in this study. The pseudoatomic calculations were performed for Ca-$3s^2$ $3p^6$ $4s^2$ and Rh-$4d^8$ $5s^1$ states in case of $CaRh_2$ and La-$5s^2$ $5p^6$ $5d^1$ $6s^2$ and Rh-$4d^8$ $5s^1$ states in case of $LaRh_2$ intermetallic. The cut-off energy of plane wave was set to 350.0 eV in this present calculation. The k-point sampling of the Brillouin zone (BZ) was employed by using the Monkhorst-Pack scheme [21] with 8×8×8 grid points in the primitive cell of both the compounds. The geometrical optimization of the crystal structure was performed by the Broyden-Fletcher-Goldfarb-Shanno (BFGS) minimization method [22]. For optimizing the crystal structure criteria of convergence were set to $1.0\times10^{-5}$ eV/atom for energy, 0.03 eV/Å for force, 0.05 GPa for stress and 0.001Å for ionic displacement. The elastic stiffness constants of the cubic laves phase $CaRh_2$ and $LaRh_2$ were obtained by the stress-strain method [23] at the optimized structure under the condition of each pressure. Then the bulk modulus was obtained using the elastic constants. In that case the criteria of convergence were set to $2.0\times10^{-6}$ eV/atom for energy, $2.0\times10^{-4}$ Å for maximum ionic displacement, $6.0\times10^{-3}$ eV/Å for maximum ionic force and 0.10 GPa for maximum stress component. The maximum strain amplitude was set to be 0.003 in the present calculation.

## III. Results and discussion

### A. Structural properties

The laves phase $CaRh_2$ and $LaRh_2$ possess face centered cubic $MgCu_2$ type structure with space group *Fd-3m* (No. 227) and their equilibrium lattice parameters are 7.52 Å and 7.64 Å respectively [12, 14]. The atomic positions of H (H = Ca and La) and Rh in the unit cell are (0, 0, 0) and (0.625, 0.625, 0.625) respectively [6]. Both the compounds have eight formula units in the unit cell. The three dimensional crystal structures with atomic arrangements of the investigated compounds and their primitive cell are shown in Fig.1. The data of investigated structural parameters are recorded in Table 1. The calculated lattice constants of $CaRh_2$ and $LaRh_2$ intermetallics are 7.65 Å and 7.82 Å respectively using GGA method and 7.45 Å and 7.63 Å respectively using LDA method. From the values of calculated lattice constants we can conclude that the experimental and theoretical values are in good agreement for both the intermetallics implying the reliability of this study. It can also be noted that LDA method is more appropriate than GGA method for the study of these two intermetallics.

**B. Single and polycrystalline elastic properties**

Elastic properties afford a fundamental role in the field of material science and modern technology. These properties provide a relation between the mechanical and dynamical activities of crystalline solids and confer essential information regarding the nature of forces working in solid materials. Basically they give information on the stability and rigidity of solid materials [24, 25]. These constants also relate the various elementary solid-state phenomena like equation of states, ductility, stiffness, brittleness, anisotropy, vibrations of normal mode and transmission of elastic waves [26]. A detailed study concerning the elastic constants and mechanical properties of Laves phase $CaRh_2$ and $LaRh_2$ are given in this part. We have determined the elastic constants from a linear fit of the calculated stress-strain function according to Hook's law [27]. For both compounds the elastic constants are calculated by GGA and LDA approximations. The calculated elastic constants are listed in Table 2. Because of cubic structures both intermetallic compounds have three independent elastic constants $C_{11}$, $C_{12}$ and $C_{44}$ fulfilling the recognized Born stability criteria [28]: $C_{11} > 0$, $C_{44} > 0$, $C_{11} - C_{12} > 0$ and $C_{11} + 2C_{12} > 0$. From Table 2 we can see that the elastic constants of $CaRh_2$ and $LaRh_2$ satisfy all of these criteria suggesting that both compounds are mechanically stable. Unfortunately, to best of our knowledge, there are no experimental and theoretical records in text for the elastic constants of Laves phase $CaRh_2$ and $LaRh_2$ available for comparison, therefore we regard as the present results as prediction study which still awaits an experimental evidence. The calculated values of bulk modulus, B from the elastic constants, by using LDA and GGA approximations have almost the same values as ones obtained from the fit to a Birche-Murnaghan EOS ($B_0$) in LDA and GGA approximations. Using the calculated values of $C_{ij}$, the most significant mechanical properties like bulk modulus $B$, shear modulus $G$, Young's modulus $E$, anisotropy factor $A$ and Poisson's ratio $v$ of Laves phase $CaRh_2$ and $LaRh_2$ are also calculated by the help of Voigt-Reuss-Hill (VRH) averaging scheme [29] listed also in Table 2. The Voigt and Reuss bounds of $B$ and $G$ for cubic systems can be represented by the following expressions [30]:

$$B_v = B_R = \frac{(C_{11} + 2C_{12})}{3} \qquad (1)$$

$$G_v = \frac{(C_{11} - C_{12} + 3C_{44})}{5} \qquad (2)$$

$$G_R = \frac{5C_{44}(C_{11} - C_{12})}{[4C_{44} + 3(C_{11} - C_{12})]} \qquad (3)$$

The arithmetic mean value of the Voigt ($B_V$, $G_V$) and the Reuss ($B_R$, $G_R$) bounds which is used to calculate the polycrystalline modulus is given by in terms of Voigt-Reuss-Hill approximations:

$$B_H = B = \frac{1}{2}(B_R + B_v) \qquad (4)$$

$$G_H = G = \frac{1}{2}(G_v + G_R) \qquad (5)$$

Using the following expressions we have also calculated the Young's modulus ($Y$) and Poisson's ratio ($v$),

$$Y = \frac{9GB}{3B + G} \qquad (6)$$

$$v = \frac{3B - 2G}{2(3B + G)} \qquad (7)$$

Young's modulus $Y$ which measures the response to a uniaxial stress averaged over all directions and is often used to indicate a measure of stiffness, *i.e.* the large value of $Y$ indicating the stiffer is the material. From Table 2 we see that both of the compounds are stiff and CaRh$_2$ is stiffer than that of LaRh$_2$.

The Zener anisotropy factor $A$ which provides a measure of degree of elastic anisotropy in solid [31] materials is obtained by using the following equation-

$$A = \frac{2C_{44}}{(C_{11} - C_{12})} \qquad (8)$$

For an isotropic crystal the value of $A$ is 1and for anisotropic crystal the values of $A$ are either smaller or greater than unity. From Table 2 we see that for both of compounds the values of $A$ are greater than unity indicating that these compounds are weakly anisotropic.

$C_{12} - C_{44}$ is defined as the Cauchy pressure and is used to illustrate the angular nature of atomic bonding [32]. For metallic compounds the value of Cauchy pressure is positive and for nonmetallic compounds the value of the Cauchy pressure is negative [33]. As shown in Table 2 both of compounds have positive value of Cauchy pressure indicating that both compounds have metallic behaviors. This result is completely satisfying the result having from the analysis of electronic band structures of these compounds discussed in section C.

The Pugh's ductility index (B/G) is one of the most extensively used malleability indicators of materials [34]. To distinguish the ductile and brittle material the critical value of B/G is 1.75. The material will behave in a brittle manner if $B/G < 1.75$ and the material perform as ductile if $B/G > 1.75$. For both of compounds the calculated values are greater than 1.75 (shown in Table 2) indicates that they behave ductile manner.

The Poisson's ratio is a useful index to comprehend the nature of bonding force in a crystal. For covalent crystal $v = 0.1$ whereas for ionic crystal $v = 0.25$. The value from 0.25 to 0.5 implies the force exists in a solid is central [35]. From Table 2, we notice that the value of $v$ for both the compounds lie between the ranges of 0.25 to 0.5 indicating the existence of central force in both intermetallics at ambient condition.

## C. Electronic properties and chemical bonding

The electronic properties such as electronic band structure, partial density of state (PDOS), total density of state (TDOS) and total charge density of $HRh_2$ have been calculated and discussed in this section. The electronic band structure of a material provides information about the material to be conductor, non-conductor and insulator. The partial and total density of state also provide information about the bonding characteristics and number of states at occupied energy level in statistical and solid state physics [36].The electronic band structure, partial density of state and total density of state are shown in Fig. 2 and Fig. 3 respectively. From band structure diagram (Fig. 2), we see that the valance bands and conduction bands overlap with each other at Fermi level which indicates that both the materials possess metallic characteristics. It can be noted from Fig. 2 that for both phases GGA and LDA approximation exhibits nearly similar result. Though, a slight variation is appeared due to different calculation method.

From Fig. 3 one can notice that in valence band the dominant feature is observed for Ca-4s and Rh-4d states for $CaRh_2$ intermetallic. In case of $LaRh_2$ La-5d and Rh-4d states dominate more in the valence band. We found similar trend between the result having from GGA and LDA approximation. The conduction band of $CaRh_2$ mostly consists of Rh-4d orbital. Though, in case of $LaRh_2$ we observe the domination of both La-5d and Rh-4d orbital in the conduction band. However at Fermi level Rh-4d state contributes more for both the materials. Though, in case of LaRh$_2$ some contribution at Fermi level comes from La-5d state. The metallic nature of both Laves phases emerges from Rh metal with slight contribution from constituent atom. The computed density of states at Fermi level for CaRh$_2$ is 5.29 states/eV and for LaRh$_2$ is 3.10 states/eV.

For further analyzing the chemical bonding properties of these two laves phase the total charge density is calculated and represented in Fig. 4. The total charge density of both the compounds is calculated for (001) crystallographic plane. From Fig. 4(a) we observe no overlapping of charge distribution between Ca and Rh atoms implying the ionic nature of Ca-Rh bond. Similar effect is observed for La-Rh bond in LaRh$_2$ as shown in Fig. 4(b). We also calculated the Mulliken atomic populations for both phase [37]. Population of Rh-Rh bond is negative for both intermetallics indicating the existence of ionic nature of this bond [38]. The ionic characteristics are a result of the metallic characteristics [39] implying the metallic nature of Rh-Rh bonds in both laves phase. Therefore, we can conclude that both the intermetallics might be explained as a mixture of ionic and metallic bonds.

**D. Optical properties**

The detailed studies of optical properties of compounds are crucial which can find potential application in semiconductor and solar cell industry. The materials which possess good optical properties have huge application in photoelectron devices. The optical properties of CaRh$_2$ and LaRh$_2$ Laves phase compounds are investigated by using the dielectric function $\varepsilon(\omega) = \varepsilon_1(\omega) + i\varepsilon_2(\omega)$ where the imaginary part of dielectric function $\varepsilon_2(\omega)$ is obtained from the momentum matrix elements between the occupied and the unoccupied electronic state by using the following equation [17];

$$\varepsilon_2(\omega) = \frac{2e^2\pi}{\Omega\varepsilon_0} \sum_{k,v,c} |\psi_k^c|u.r|\psi_k^v|^2 \, \delta(E_k^c - E_k^v - E) \tag{9}$$

Where, $u$ is a vector which defining the polarization of incident electric field, $e$ is the electronic charge, $\omega$ is the light frequency, $\psi_k^c$ and $\psi_k^v$ are the conduction and valance band wave function. The real part $\varepsilon_1(\omega)$ of the dielectric function can be obtained from the imaginary part by using the Kramers-Kronig transform. The other optical properties, such as refractive index, conductivity, optical reflectivity, absorption coefficient and loss-function are also obtained from the value of $\varepsilon(\omega)$ by using the equations given in ref. [17].

The calculated optical functions of $HRh_2$ as a function of photon energy up to 40 eV toward the [100] and [001] polarization vector using both GGA and LDA approximation are illustrated in Fig. 5 and Fig. 6. Fig. 5(a) and Fig. 6(a) illustrates the reflectivity spectra of $CaRh_2$ and $LaRh_2$ intermetallics as a function of photon energy. One can see that the reflectivity is 0.70-0.65 in the infrared region (1.24 meV-1.7 eV) for $CaRh_2$ and 0.69-0.64 for $LaRh_2$ phase. Reflectivity increases in the visible part of the spectrum and then drops sharply and becomes zero in the ultraviolet region with some peaks for both the compounds. The maximum value of the reflectivity is 0.97 for $CaRh_2$ and exactly 1.0 for $LaRh_2$ compound. Fig. 5(b) and Fig. 6(b) illustrates the absorption coefficient of $CaRh_2$ and $LaRh_2$ intermetallics. Absorption coefficient implies how far light of particular energy can enter into a compound before being completely absorbed. We observe three peak with the highest peak located at 27 eV for $CaRh_2$ and 19 eV in case of $LaRh_2$ compound implying good absorption coefficient in the ultraviolet region. The peaks shift to high energy region when La is replaced by Ca in $HRh_2$ structure. The real part of refractive index is shown in Fig. 5(c) and Fig. 6(c) for $CaRh_2$ and $LaRh_2$ intermetallics respectively. It denotes the phase velocity of wave. For both compounds the value of $n$ decreases with the increase in photon energy. Fig. 5(d) and Fig. 6(d) represents the imaginary part of refractive index of $CaRh_2$ and $LaRh_2$ intermetallics. It denotes the absorption loss of electromagnetic wave when propagates into a medium. Except some variation in the peak position we found similar trend between the spectrums of both the phase. The real and imaginary part of the dielectric function for $CaRh_2$ and $LaRh_2$ Laves phase compounds are shown in Fig. 5(e), Fig. 5(f) and Fig. 6(e), Fig. 6(f) respectively. The response of a compound to the incident electromagnetic wave is characterized by the dielectric function. We observe no peak in the real part of dielectric function for both the compounds whereas some minor peaks are found in the imaginary part of the dielectric function. The value of imaginary dielectric constant $\varepsilon_2$ becomes zero at about 9 eV and 10 eV for $CaRh_2$ and $LaRh_2$ intermetallics respectively. These values indicate that the corresponding compound will be transparent above the corresponding energy values [40]. Non zero value of $\varepsilon_2$ from 0-9 eV for $CaRh_2$ and 0-10 eV for $LaRh_2$ implying that absorption occurs in these energy region. The static value of dielectric constant for $CaRh_2$ and $LaRh_2$ intermetallics are 58 and 35 respectively indicating that $CaRh_2$ is a good dielectric material than $LaRh_2$. Fig. 5(g) and Fig. 6(g) illustrates the conductivity spectra of $CaRh_2$ and $LaRh_2$ intermetallics as a function of photon energy. For both the compounds photoconductivity starts from zero photon energy implying the metallic nature of $CaRh_2$ and $LaRh_2$ intermetallics. Fig. 5(h) and Fig. 6(h) represents the energy loss spectra of $CaRh_2$ and $LaRh_2$ intermetallics. We observe single peak at around 12 eV for $CaRh_2$ and 13 eV for $LaRh_2$ intermetallic. Though, GGA and LDA results show slight variation. These peaks are related to the rapid demission in the reflectance of the corresponding compound.

## IV. Conclusions

In summary, in this letter first-principle simulations based on DFT have been used to predict the detailed physical properties of two Laves phase compounds $CaRh_2$ and $LaRh_2$. The present study predicts the metallic nature of these two intermetallics. The study of elastic constants ensures the mechanical stability of the two phases. Both the compounds are stiff and $CaRh_2$ is stiffer than that of $LaRh_2$. The value of Zener anisotropy factor shows that both the intermetallics are anisotropic in nature. The analysis of Pugh's ratio shows the ductile nature of both the phases. The study of DOS exhibits that most of the contribution originates from Rh-4d states at Fermi level. The study of bonding characteristics reveals the existence of ionic and metallic bonds in both intermetallics. Both the compounds exhibit rather good optical characteristics. We expect the present study will help to consider these two Laves phase compounds for many technological applications in future.

**Acknowledgement**

We would like to thank Department of Physics, Pabna University of Science and Technology, Bangladesh, for the laboratory support.

**Table 1.** Structural parameters of $CaRh_2$ and $LaRh_2$ intermetallics.

| Properties | $CaRh_2$ | | $LaRh_2$ | |
|---|---|---|---|---|
| | This study | Expt. [12] | This study | Expt. [14] |
| $a_0$ (Å) | 7.65[a], 7.45[b] | 7.525 | 7.82[a], 7.63[b] | 7.64 |
| $V_0$ (Å$^3$) | 447.69[a], 413.49[b] | 426.10 | 478.21[a], 444.19[b] | 445.94 |
| $B_0$ (GPa) | 120.11[a], 135.34[b] | - | 122.54[a], 160.32[b] | - |

(GGA)[a], (LDA)[b]

**Table 2.** The independent elastic constants $C_{ij}$ (GPa), Cauchy pressure ($C_{12} - C_{44}$), bulk modulus $B$ (GPa), shear modulus $G$ (GPa), Young's modulus $Y$ (GPa), $B/G$ values, Poisson's ratio $v$ and anisotropy factor $A$ of Laves phase $CaRh_2$ and $LaRh_2$.

| Compounds | | $C_{11}$ | $C_{12}$ | $C_{44}$ | $C_{12}$ - $C_{44}$ | $B$ | $G$ | $Y$ | $B/G$ | $v$ | $A$ |
|---|---|---|---|---|---|---|---|---|---|---|---|
| $CaRh_2$ | GGA | 175 | 99 | 64 | 35 | 124.33 | 51.93 | 135.75 | 2.39 | 0.317 | 1.68 |
| | LDA | 209 | 121 | 76 | 45 | 150.33 | 61.04 | 161.29 | 2.46 | 0.321 | 1.727 |
| $LaRh_2$ | GGA | 157 | 93 | 65 | 28 | 114.33 | 48.91 | 128.42 | 2.34 | 0.313 | 1.383 |
| | LDA | 206 | 127 | 73 | 54 | 153.33 | 57.21 | 152.65 | 2.68 | 0.334 | 1.484 |

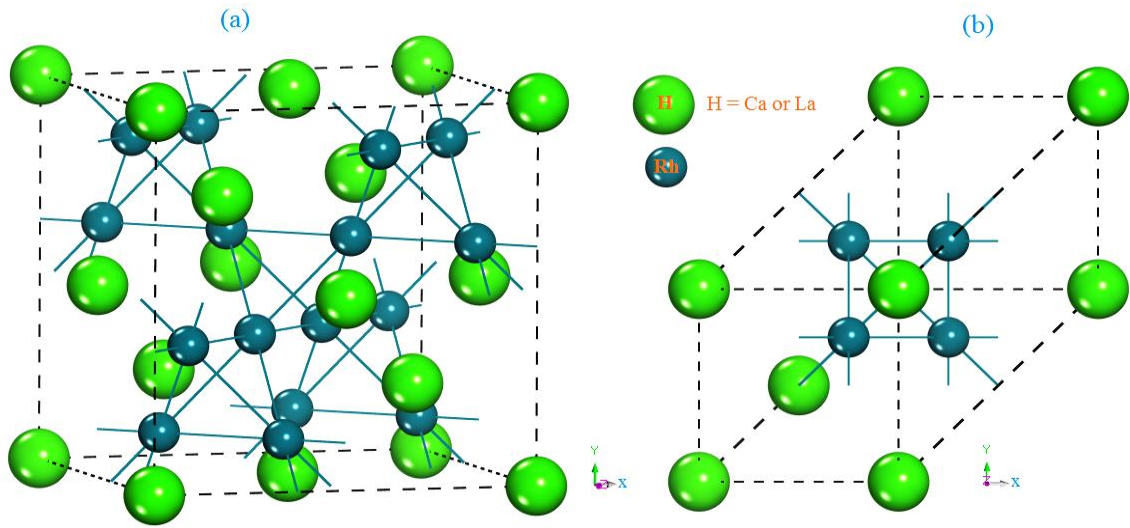

**Fig. 1.** Crystal structures of $HRh_2$ (H = Ca or La). (a) conventional unit cell and (b) primitive cell.

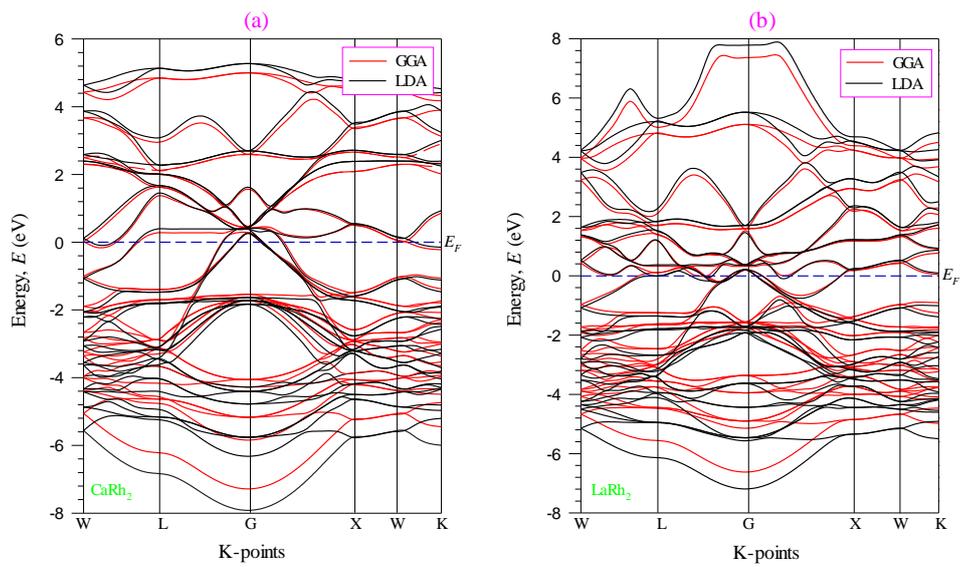

**Fig. 2.** Electronic band structures of $HRh_2$ (H = Ca or La).

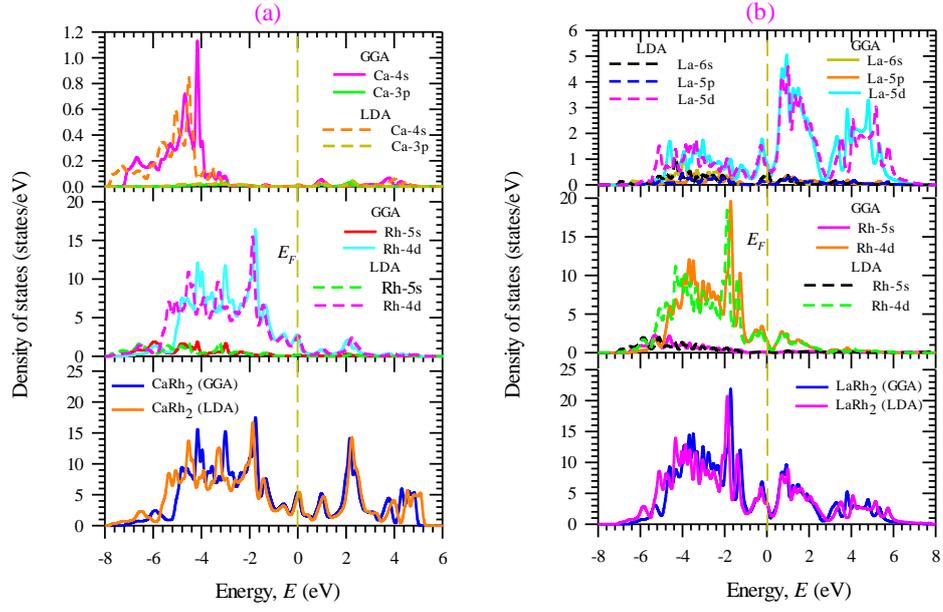

**Fig. 3.** Total and partial density of states of HRh$_2$ (H = Ca or La).

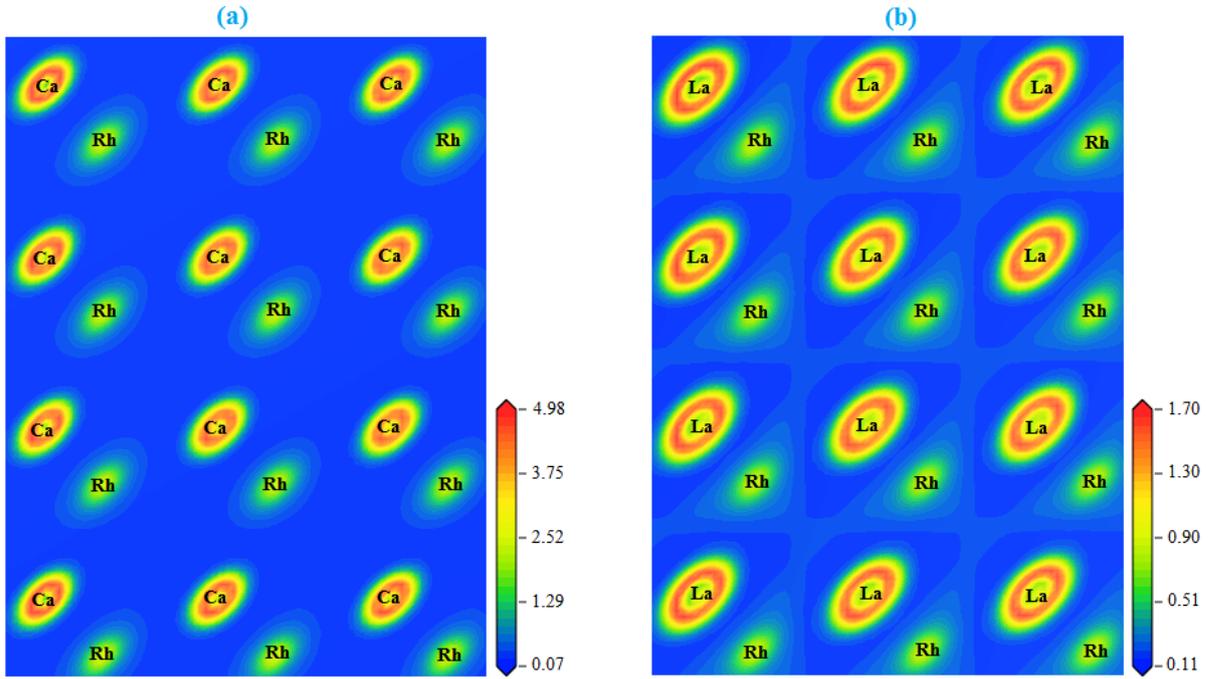

**Fig. 4.** Total charge density of HRh$_2$ (H = Ca or La). (a) for CaRh$_2$ (b) for LaRh$_2$

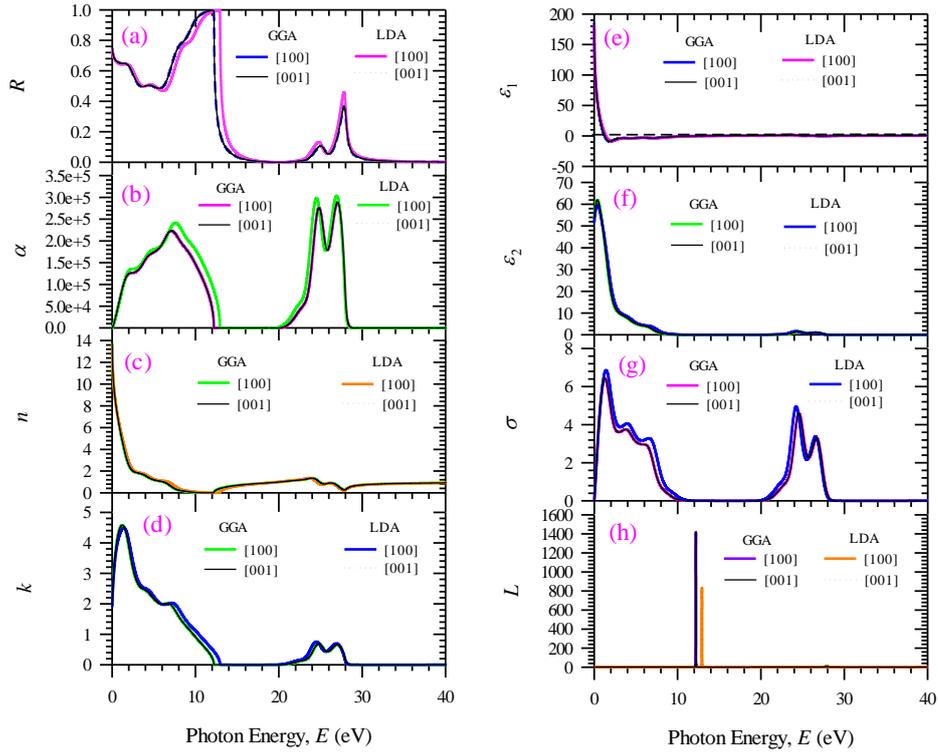

**Fig. 5.** The optical functions (a) reflectivity, (b) absorption coefficient, (c) real part of refractive index, (d) imaginary part of refractive index, (e) real part of dielectric function, (f) imaginary part of dielectric function (g) conductivity, and (h) loss function of $CaRh_2$ Laves phase for polarization vector [100] and [001].

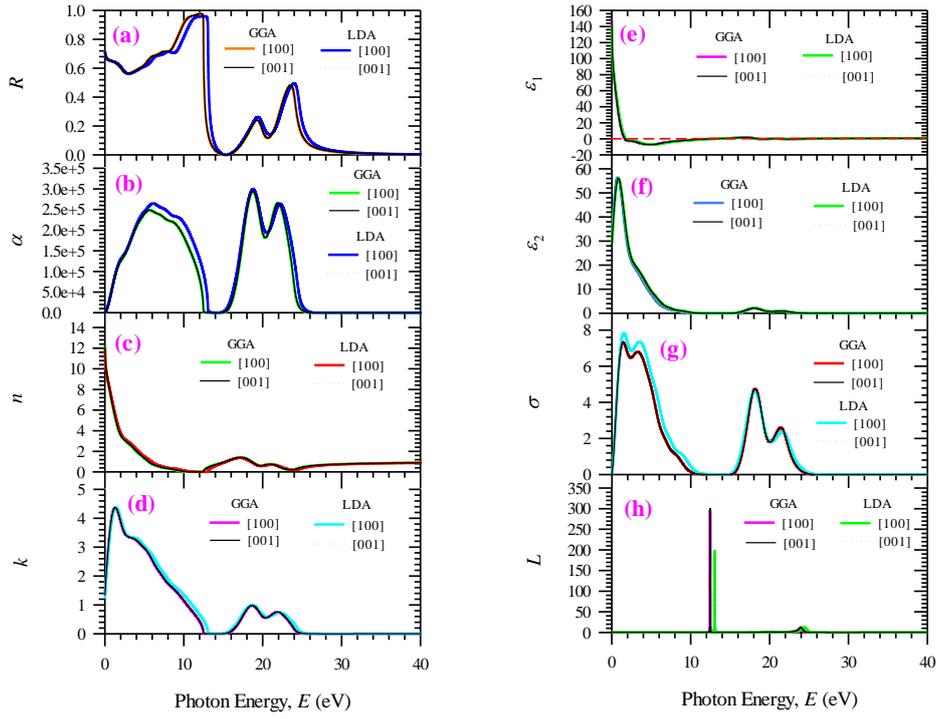

**Fig. 6.** The optical functions (a) reflectivity, (b) absorption coefficient, (c) real part of refractive index, (d) imaginary part of refractive index, (e) real part of dielectric function, (f) imaginary part of dielectric function (g) conductivity, and (h) loss function of LaRh$_2$ Laves phase for polarization vector [100] and [001].